\documentstyle[preprint,eqsecnum,aps]{revtex}

\begin{document}
\input epsf.sty
\draft
\preprint{HEP/123-qed}
\title{Probabilistic Approach to Time-Dependent
Load-Transfer Models of Fracture
}
\author{J.B.G\'omez}
\address{Departamento de Ciencias de la Tierra.
 Universidad de Zaragoza. 50009. Zaragoza (Spain)
}
\author{Y.Moreno\renewcommand{\thefootnote}{\fnsymbol{footnote}}\footnote[1]
{On leave from Department of Physics, Technological University of
Havana (ISPJAE), Havana 19390, Cuba.}
\renewcommand{\thefootnote}{\arabic{footnote}} and A.F.Pacheco}
\address{Departamento de F\'{\i}sica Te\'orica. Universidad de
Zaragoza. 50009. Zaragoza (Spain).
}
\date{\today}
\maketitle
\begin{abstract}
A probabilistic method for solving time-dependent load-transfer
models of fracture is developed. It is applicable to any rule of load
redistribution, i.e, local, hierarchical, etc. In the new method, the
fluctuations are generated during the breaking process (annealed randomness)
while in the usual method, the random lifetimes are fixed at the
beginning (quenched disorder). Both approaches are equivalent.
\end{abstract}

\narrowtext

\section{introduction}

The modelization of fracture in disordered systems is a subject of
great interest in natural and artificial materials \cite{herrman90}. A time-dependent method to
describe the failure of
materials under stress, within the fiber-bundle paradigm, was proposed
by Coleman \cite{coleman57}. In this model a set (bundle) of elements
 (fibers)
is considered with each element having a prescribed lifetime when
subject to an applied stress (load). When elements fail, their load
is redistributed to other elements of the set according to a
prescribed rule of transfer. As a consequence of the load transfer,
the lifetime of the receptors is actually reduced and the question is: how long
does it take for the whole set to collapse?
These fiber-bundle models are called dynamical, or time-dependent
\cite{coleman57,smith81,phoenix83},
as opposed to their static counterparts, which have also been
intensively studied \cite{daniels45,harlow91,turcotte82}.
 The rule for redistributing the load of failed elements can be
 wide, but there are two limiting cases. In the first, the
stress of the failed element is transferred equally to all surviving
 elements (ELS, for Equal Load Sharing). In the second, the load of the
 failed element is transferred to the nearest surviving element(s)
(LLS, for Local Load Sharing). Hierarchically organized transfer
 (HLS) criteria are also of great interest \cite{turcotte82,smalley85,newman91}.
 Recently, these models have received much attention in the
geophysical literature \cite{Turcotte97}, because one would reasonably
expect the emergence of universal
 scaling laws of the type observed in seismology \cite{Newman94,newman95}. In this field, the
bundle is a representation of a fault,
and the individual elements or fibers represent asperities on the fault plane.

In Reference\cite{newman95}, the ELS case is formulated in terms of a differential equation of
the radioactive decay type. We
 have followed this perspective to devise a numerical probabilistic method to deal with any type of
transfer
rule. In Section \ref{sec:themethod}, we explain in detail the differences between
the usual approach and the new probabilistic approach. In Section \ref{sec:results}, we
compare results and present a brief discussion.

\section{the methods}
\label{sec:themethod}

Suppose a set of \(N_{0}\) elements identified
 on the sites of a supporting lattice. This information is contained in a list
 \(\{\vec{x}_{i,t}\}\equiv\{\vec{x}\}_{t}\) , \(1\le i\le N_0\). This list is
necessary, except in the ELS case, to know how to distribute the load of the
failed elements. The broken elements are marked in \(\{\vec{x}\}_{t}\). At \(t=0\), all the
elements of the set are loaded with a reference value \(\sigma_{0}=1 \). At any time, the total
load acting on the surviving elements is constant, equal to \(N_0\sigma_0\).

To each element, \(i\), one assigns a random lifetime, \(t_{i,0}\), under the
unity of stress:
\begin{equation}
\ n_i=1-e^{-t_{i,0}}\label{one},
\end{equation}
where \(n_i\) are random numbers between 0 and 1. This choice implies a logarithmic distribution
of lifetimes. A more general distribution function for the failure time of a single element
subjected to a known load history \(\sigma(t)\) is (see, for example \cite{phoenix83,Newman94})
\begin{equation}
\ n_i=1-exp\left[ -\Psi\left( \int_0^{t_{i,0}} \nu [\sigma(\tau)]d\tau \right)
\right]\label{twotwo},
\end{equation}
 where \(\Psi(x)\) is the {\it shape function\/}. The time integral in Eq.\ (\ref{twotwo})
introduces a hazard rate \(\nu(\sigma)\) known generally as the {\it breakdown rule\/} in terms
of the instantaneous load level. Experimental and theoretical work \cite{coleman57,phoenix83}
favors a shape function \(\Psi(x)\) of the form
\begin{equation}
\ \Psi(x)=x^\beta\label{twothree} ,
\end{equation}
known as the Weibull shape function, with the particular choice \(\beta=1\) giving the
exponential shape function.

As for the breakdown rule, two special forms are widely used in the literature: the exponential
breakdown rule,
\begin{equation}
\ \nu_{e}=\phi e^{\eta (\sigma/\sigma_{0})}\label{twofour} ,
\end{equation}
 and the power-law breakdown rule,
\begin{equation}
\ \nu_{p}=\nu_{0}\left( \frac{\sigma}{\sigma_{0}} \right)^{\rho}\label{twofive},
\end{equation}
 with \(\phi\), \(\eta\), \(\rho\), \(\nu_{0}\), \(\sigma_{0}\) all positive constants. \(\rho \)
is called the Weibull exponent because inserting Eq.\ (\ref{twothree})
and Eq.\ (\ref{twofive}) in Eq.\ (\ref{twotwo}) mimics the static
Weibull distribution for the failure load of a single element. This
parameter typically varies between 2 and 5. The exponential breakdown
rule, Eq.\ (\ref{twofour}) has a characteristic failure rate, whereas
the power-law breakdown rule is scale free and can be regarded as a
local approximation to the former. Following \cite{newman95}, we will
use Eq.\ (\ref{twofive}) as the individual breakdown rule in order to
be able to compare the perfomance of the two approaches for
load-transfer models. For further insights into the theoretical and
experimental basis of the Weibull shape function see \cite{coleman57},
and for the breakdown rules see References \cite{phoenix83} and
\cite{Newman94}.

Without losing generality one can choose \(\nu_{0}=\sigma_{0}=1\) in
Eq.\ (\ref{twofive}), which means that \(\sigma^{\rho}\) is a measure
of the failure rate (\(i.e\), a unit failure rate under the unity of
load is assumed). As \(\nu_{0}\) is actually a frequency,
\(\nu_{0}t\) is a dimensionless time variable and because of the
particular choice \(\nu_{0}=1\), \(t\) will hereafter stand for
non-dimensional time.

If one substitutes Eq.\ (\ref{twothree})and Eq.\ (\ref{twofive}) in
Eq.\ (\ref{twotwo}) with the particular set of constants
\(\beta=\nu_{0}=\sigma_{0}=1\), we obtain
\begin{equation}
\ n_i=1-exp\left(- \int_0^{t_{i,0}} \sigma^{\rho}(\tau)d\tau \right),
\end{equation}
 which can be integrated for constant unit load \(\sigma(t)=\sigma_{0}=1\) to give Eq.\
(\ref{one}).

 When loads of failed elements are redistributed, the load acting
 on each element will no longer
 be the constant \(\sigma_{0}\) but will depend on time
\(\sigma_{i}(t)\ge\sigma_{0}=1 \). Thus we introduce a reduced time to
failure for each element, \(T_{i,f}\), given by
\begin{equation}
\ t_{i,0}=\int_0^{T_{i,f}}\left[ \frac{\sigma_{i}(t)}{\sigma_{0}} \right]^{\rho}dt\label{two}.
\end{equation}
 In the case of independent elements, \(\sigma_{i}(t)=\sigma_{0}=1 \) and \(t_{i,0}=T_{i,f} \).
 However, load transfer occurs, and hence the actual time to failure of element
  \(i\), \(T_{i,f}\) is reduced to below
 \(t_{i,0}\). By imposing the fulfillment of Eq.\ (\ref{two}), the successive
order of breaking of the \(N_{0}\) elements, one after the other, is easily identified
 and the total time of collapse is the \(T_{i,f}\) of the longest lasting element. Thus, in this
approach the randomness, that  is the population of lifetimes, is
fixed at
\(t=0\) (quenched disorder), and the breaking process is completely
deterministic. Henceforth, we will refer to this approach as the usual
one.

In the new probabilistic approach presented here, the fluctuations are
generated during the breaking process and hence it is an example of
so-called annealed randomness. An interesting question is whether the
two
 types of disorder, namely quenched and annealed disorder, in these
models lead to different results, as has been observed for some
critical phenomena \cite{herrman90}.

In Ref. \cite{newman95}, Newman {\sl et al\/} formulated the ELS mode in terms of a
differential equation of the radioactive decay type.
Denoting the number of surviving elements as \(N_{s}(t)\), its
differential change is given by
\begin{equation}
\frac{dN_{s}}{dt}=-N_{s}\sigma^\rho\ \label{tres},
\end{equation}
hence \(\sigma^\rho\) represents the decay rate. But in the ELS mode
\(\sigma=\left(\frac{N_{0}}{N_{s}}\right)\), hence
\begin{equation}
N_{s}(t)=N_{0}[\rho(T_{f}-t)]^{1/\rho}\ \label{eight}
\end{equation}
and \(T_{f}=1/\rho\). In this setting, fluctuations do not exist
 and one simply obtains mean values for the failure rate.

Following the perspective of the previous differential equation in which a
group of elements, supporting the same load \(\sigma\), fail at a rate
\(\sigma^\rho\) or, in other words, have a mean-life \(1/\sigma^\rho\), one
can devise a probabilistic method for any transfer rule.
 The scenario would be a set of \(N_{0}\) elements identified
 in \(\{\vec{x}\}_{t}\), like a sample of radioactive nuclei fixed on a lattice, all having
 initially a decay rate \(\sigma_{0}^{\rho}=1\). As time passes, failures (disintegrations)
 occur and this does not merely imply the effective disappearance of the failed
elements, but also the modification of the decay rate of other
surviving elements. The modification comes from the redistribution of
load as accorded in the rule of transfer (ELS, LLS, etc), and the
assumption that the decay rate of any element is given by its
\(\sigma^\rho\) value. As in this strategy of calculation one has to proceed at
discrete time intervals, \(\delta_{j}\), \(j=1,2,...\), the
information of loads in the set will be contained in a list denoted by
 \(\{\sigma_{i,j}\}\equiv\{\sigma\}_{j}\). The list is updated at each time
step, together with \(\{\vec{x}\}_{j}\). After \(j\) time steps, there will
have appeared subsets of elements. Each subset is formed by all the surviving elements bearing
the same load. We organise these subsets into sublists identified by the subindex \(l\),
and denote the corresponding load by \(Y_{l}\) and the number of elements belonging to the
sublist \(l\) by \(N_{l}\).
 This information, which is obtained from \(\{\sigma\}_{j}\), will be denoted by
\(\{Y_{l},N_{l}\}_{j}\) and updated simultaneously. At the beginning, as
 the load of all elements is 1, the sublists are

\(Y_{l,0}=1, N_{l,0}=N_{0}\) if \(l=1\),

\(Y_{l,0}=0, N_{l,0}=0\) if \(l\not=1\).

Now, it is clear that the simultaneous existence of several sublists in the sample,
each with a different decay rate \(Y_{l}^\rho\), poses a difficulty for an accurate
description of the decay process of the whole set \cite{trece}.  The key point is the
choice of the length of the time intervals, \(\delta_{j}\). To illustrate this problem, in Fig.\
\ref{figure1} we have plotted the detailed evolution of breaking of a hierarchical set of
\(N_{0}=1024\), coordination number equal to 2, and \(\rho=4\). In abscisas one represents time,
from \(0\) to \(T_{f}\). In ordinates the spatial position of the
\(N_{0}\) elements of the set is represented. At \(t=0\) all elements
are sound. As time evolves, breakings (represented by small crosses)
are produced and therefore the number of failed elements, represented
by the continuous line, grows. At \(t=T_{f}\) the number of failures
is
\(N_{0}\). The height of the vertical spikes represents the load supported by an
element at the time of failing. For short times, ruptures appear
dispersed across the set and the rate of breaking is small.
Progressively though, there appear cracks formed by the failures of
neighboring elements, and this makes the continouos line adopt steeper
slopes. Finally, the final breakdown occurs related to a big crack of
a size similar to the whole system. This stage is also related to the
high values of the spikes. The progressive acceleration of the
breaking process is thus clear from this figure.

Therefore the time interval used in the probabilistic approach must be
variable with time. Otherwise, if one takes for
\(\delta\) a reasonable value for the beginning, the final part of the
breaking will be badly described: in each \(\delta\), many elements
will fail and the prediction of \(T_{f}\) will be very inaccurate. On
the other hand, if one chooses a \(\delta\) typical of the final
stages, it will be so short that although the calculation would be
extremely good, at the beginning one would become bored awaiting the
outcome of a breaking event. That is to say, these small intervals are
not realistic for practical use.

It is for this reason that when one tries to devise an efficient
numerical method to accurately describe the time evolution of the
system, the choice of the time interval must be adjusted to the
characteristic scale at which individual elements break in the
process. This characteristic time scale, \(\delta\), as mentioned
before changes with time, \(\delta_{j}\), and is implemented through
the following definition
\begin{equation}
\delta_{j}=\quad\mbox{minimun of}\left\{\frac{1}{N_{l,j}Y_{l,j}^{\rho}}\frac{1}{\nu}\right\}\
\label{nine}
\end{equation}
 where \(\nu\) is a constant \(\ge1\), independent of \(l\) and of \(j\); we will
call it the time resolution parameter. The length of \(\delta_{j}\) as defined in
 Eq.\ (\ref{nine}) points, at each \(j\), to a specific sublist whose \(l\) will
be denoted as \(k_{j}\). Now we define a probability of failure for each sublist,
\begin{equation}
p_{l,j}=Y_{l,j}^{\rho}\delta_{j}\ \label{ten} .
\end{equation}

As \(Y_{l,j}^{\rho}\) is the failure rate for elements in sublist \(l\),
\(Y_{l,j}^{\rho}\delta_{j}\) is the expected number of casualties per element in sublist \(l\),
\(i.e\), the probability of failure for sublist \(l\). The product \(N_{l,j}\ p_{l,j}\) is
maximum for the sublist \(k_{j}\), in this case  \(N_{k,j}\
p_{k,j}=1/\nu\), which means that when the comparisons below\
(\ref{allequations}) are perfomed, elements belonging to \(k_{j}\) are
the most likely to fail. In particular if \(\nu=1\), one element of
the \(k_{th}\) sublist is likely to fail. For the other sublists, the
probability of an individual failure is lower than one. However, any
element of any \(l_{j}\) has a non-zero chance of failing in this
probabilistic approach. We have called \(\nu\) the time resolution
parameter because if it grows the time intervals
 \(\delta_{j}\) are smaller and therefore it is obvious that the process of failure is more
finely
resolved. Then the probability \(p_{l,j}\) is compared, for each element belonging to the sublist
\(l\), with a random number, \(n\), \(0\le n\le 1\).

\begin{mathletters}
\label{allequations}
\begin{equation}
\mbox{If}\quad p_{l,j}>n \quad\mbox{ the element fails.}\ \label{equationa}
\end{equation}
\begin{equation}
\mbox{If}\quad p_{l,j}<n \quad\mbox{ the element survives.}\ \label{equationb}
\end{equation}
\end{mathletters}

The elements that fail in any of the sublists
 transfer their load according to the rule
 of transfer and the information contained in the list \(\{\vec{x}\}_{j}\).
In the case that no element fails,
a new time interval \(\delta_{j+1}\) equal to \(\delta_{j}\) is added and the same probabilities,
\(p_{l,j+1}=p_{l,j}\), are compared with random numbers. This is repeated until at
least one failure occurs which modifies \(\{\vec{x}\}_{j}\), \(\{\sigma\}_{j}\),
\(\{Y_{l},N_{l}\}_{j}\),
 and hence \(\delta_{j}\). The total time to failure, \(T_{f}\), is the sum of the \(\delta_{j}\)
up
to the disappearance of all the elements.

In the ELS case, \(\delta_{j}\) can be explicitly written. After \((j-1)\) steps and assuming one
failure per step, the number of surviving elements forming the unique sublist is \(N_{j}=N_{0}-
(j-1)\), and the individual load is \(Y_{l}=(N_{0}/N_{j})\). Then
\begin{equation}
\delta_{j}=\frac{1}{N_{0}-(j+1)}\left(\frac{N_{0}-(j-1)}{N_{0}}\right)^{\rho}=\frac{(N_{0}+1-
j)^{\rho-1}}{N_{0}^{\rho}} \label{tenone} .
\end{equation}
Note that we have used \(\nu=1\) to be in accordance with the one
failure per step assumption. In Eq.\ (\ref{tenone}) one observes that,
in the first step, \(\delta_{1}=1/N_{0}\), and in the last step,
\(\delta_{N_{0}}=1/N_{0}^{\rho}\). Now we proceed to sum up all the
time intervals. In the continouos limit, we find
\begin{equation}
T_{f}=\int_{1}^{N_{0}}\frac{(N_{0}+1-x)^{\rho-1}}{N_{0}^{\rho}}dx=\frac{1}{\rho}\left[1-
\frac{1}{N_{0}^{\rho}}\right]
\end{equation}
which tends to the correct result \(1/\rho\) in the limit of large
\(N_{0}\).

\section{results and discussion}
\label{sec:results}

For the ELS case, \(\rho=2\), \(N_{0}=100\), we plot in Fig.\ \ref{figure2} the average of
\(T_{f}\) after the number of simulations expressed in the abscisas, for various \(\nu\).
 The horizontal lines comprise the extremes of the  values obtained in 10 averages of 32000 simulations
 each by means of the usual method. One can observe, a) the actual \(T_{f}\) of this set
is not \(\frac{1}{\rho}\) as predicted by the differential equation,
this being a finite-size effect; and b) \(\nu=4\) is already
sufficient in this method to reproduce the result of the usual
approach.
 For the HLS case, \(\rho=2\) and coordination number of the Cayley tree, \(c=2\),
we show in Fig.\ \ref{figure3} the dispersion of \(T_{f}\) emerging from the usual method
(squares) and from
 the new method (circles) with \(\nu=1\). Note the slight shift rightwards of
the center of the Gaussian, i.e, the values are longer. This is in agreement with what is seen
in Fig.\ \ref{figure2}. A greater value of \(\nu\) would move the Gaussian to the left up to
the coincidence.
 In Fig.\ \ref{figure4}, the averaged rates of breaking of a set of \(N_{0}=128\) are plotted
under the
 HLS rule, \(c=2\), for two values of the Weibull exponent \(\rho=4\) and \(\rho=6\). We compare
the habitual method and
the new method for \(\nu=1\).
 In Table\ \ref{table1}, a set of values of \(T_{f}\) and their intrinsic width is shown
for the HLS case, \(c=2\), by varying \(N_{0}\) and \(\nu\).
 Data for \(N_{0}=128\), and \(N_{0}=512\) are averages over 32000, and 10000 realizations
respectively. The errors quoted are one standard deviation of the mean.
 Perhaps the most abrupt rule of transfer that one can imagine is that of the
local one-dimensional unilateral model \cite{Gomez94}, where the load of failed
elements is transferred to the nearest neighbor in the row going in one direction. This implies
the almost immediate opening of big cracks and hence a great instability. The probabilistic
approach for this model has been tested and again both methods coincide.

Note that in the probabilistic method, there can be \(\delta_{j}\) in
which no element fails, and others in which several elements do.
 In contrast to the usual method, here no disorder is fixed at the start: we
 begin with \(N_{0}\) elements, all with the same mean-life; the random successive failures
are responsible for the fluctuations, i.e, this is an example of annealed disorder.
  For a small value of \(\nu\), the results emerging from the probabilistic
method are already indistinguishable from those deriving from the
usual method. So, we have numerically proved the equivalence of the
two approaches. If \(\nu\) is greater, the method demands more effort
but the results reach a saturation point. Comparing the respective
disadvantanges of computing: in the probabilistic approach it is
necessary to deal with larger sets of random numbers while in the
usual method the set of stored data is much bigger.

We conclude by quoting Feynman who, in his original paper on path integral
formalism \cite{feynman48}, writes ``although it does not yield new results
there is a pleasure in recognizing old things from a new point of view''.

\section*{acknowledgments}

This work was supported in part by the Spanish DGICYT, by the grant PB93-0378. Y.M. thanks
the AECI for a doctoral fellowship.

\begin{figure}
\caption{Rate of breaking (continouos line) vs. time of a hierarchical set. Small crosses
represent the position of local fractures. The height of the vertical
spikes indicates the load of an element at the time of its failure.
Read the text for details.}
\label{figure1}
\end{figure}
\begin{figure}
\caption{\(T_f\) results using probabilistic method for various \(\nu\).
 The two horizontal lines show the prediction of the usual method.}
\label{figure2}
\end{figure}
\begin{figure}
\caption{Comparison of the two methods, for a hierarchical model.}
\label{figure3}
\end{figure}
\begin{figure}
\caption{Evolution of \(N_s\) over time for both methods.}
\label{figure4}
\end{figure}

\begin{table}
\caption{Comparison of the two methods (HLS, c=2)(see text for details).}
\begin{tabular}{ddddd}
 &\multicolumn{2}{c}{$\rho=2$}&\multicolumn{2}{c}{$\rho=4$}\\
Method&$N_0$=128\tablenotemark[1]&$N_0$=512\tablenotemark[2]&$N_0$=128&$N_0$=512\\
\tableline
Prob$(\nu=1)$&0.3890&0.3614&0.0992&0.0750\\
Prob$(\nu=4)$&0.3807&0.3577&0.0907&0.0718\\
Prob$(\nu=10)$&0.3781&0.3573&0.0901&0.0710\\
Usual&0.3776&0.3565&0.0888&0.0705\\
\end{tabular}
\tablenotetext[1]{Simulations with $N_0$=128 elements are averages over 32000 realizations.
Standard deviation of the mean value is $\pm$2 units in the least significant digit.}
\tablenotetext[2]{Simulations with $N_0$=512 elements are averages over 10000 realizations.
Standard deviation of the mean value is $\pm$1 unit in the least significant digit.}
\label{table1}
\end{table}

\newpage

\begin{figure}
\epsfxsize=8.6cm
\epsfbox{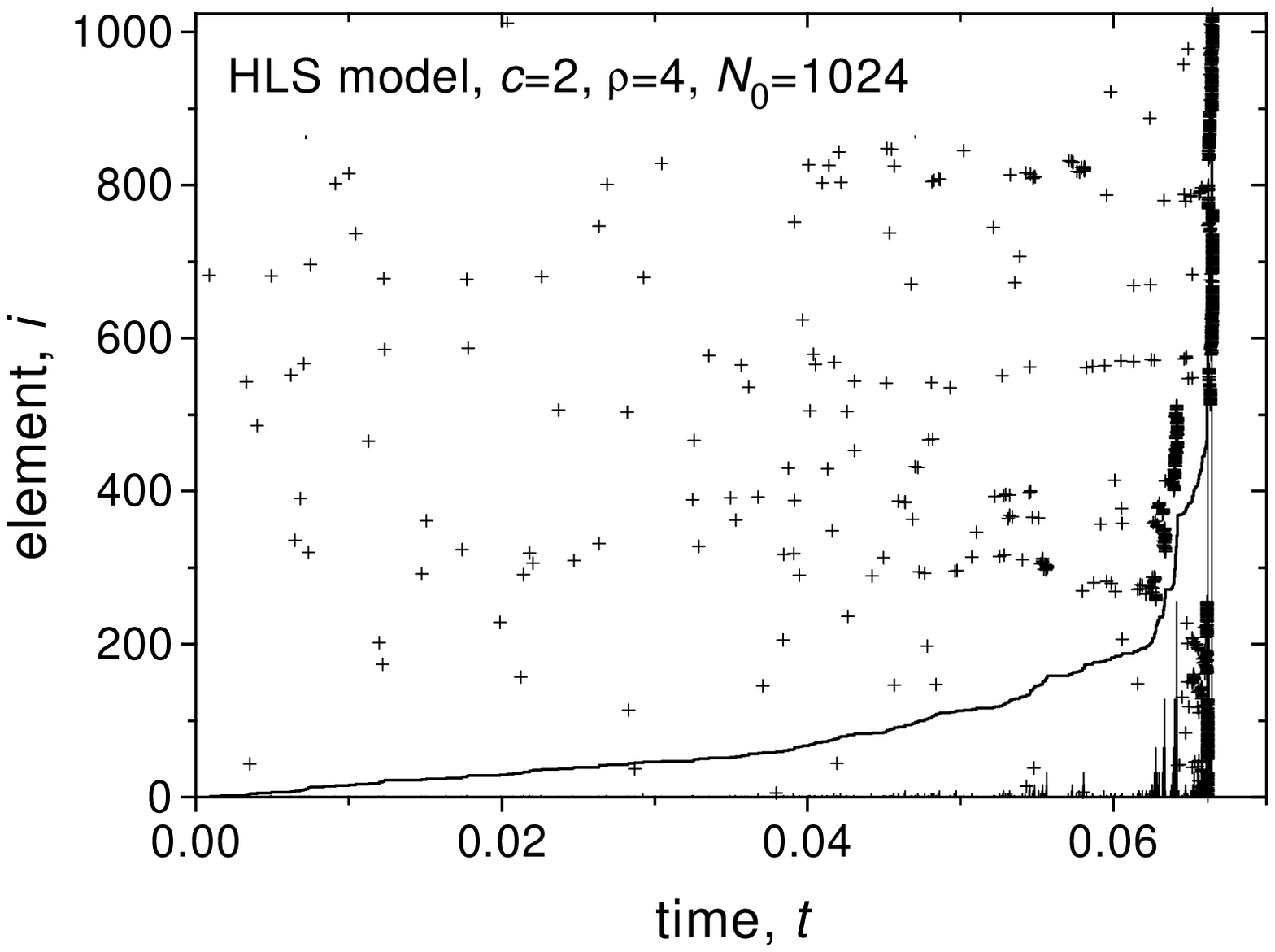}
\end{figure}

\newpage

\begin{figure}
\epsfxsize=8.6cm
\epsfbox{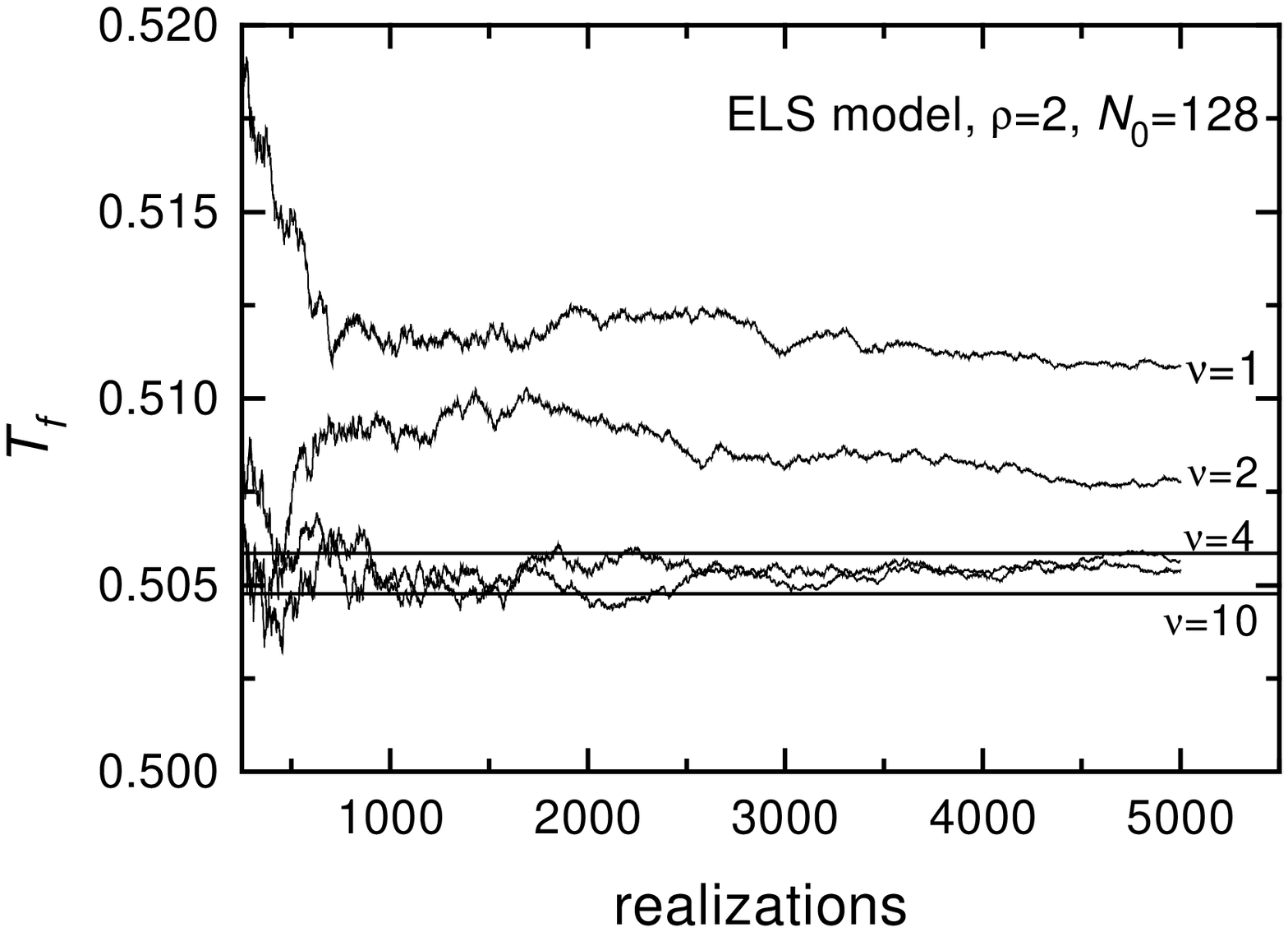}
\end{figure}

\newpage

\begin{figure}
\epsfxsize=8.6cm
\epsfbox{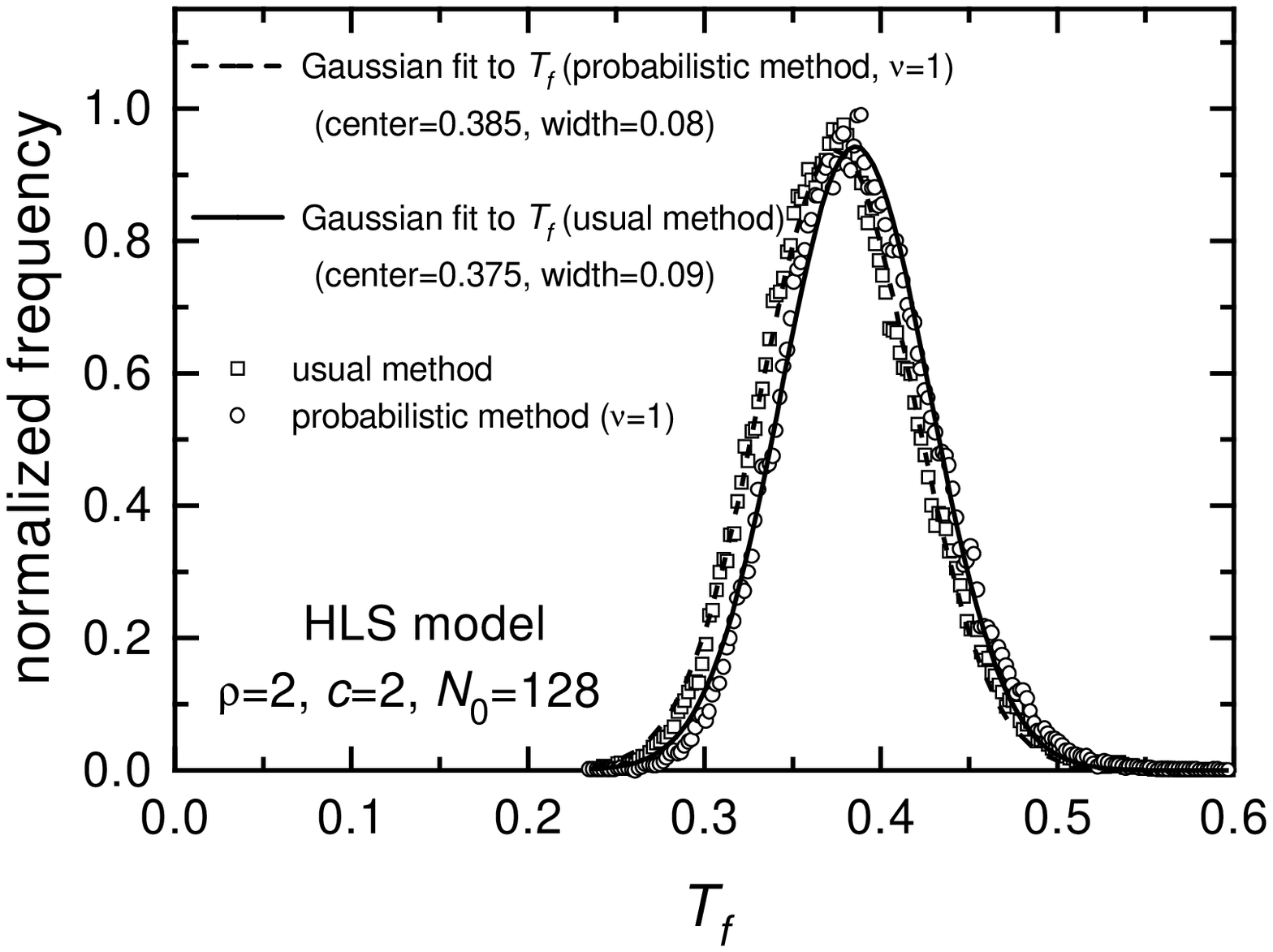}
\end{figure}

\newpage

\begin{figure}
\epsfxsize=8.6cm
\epsfbox{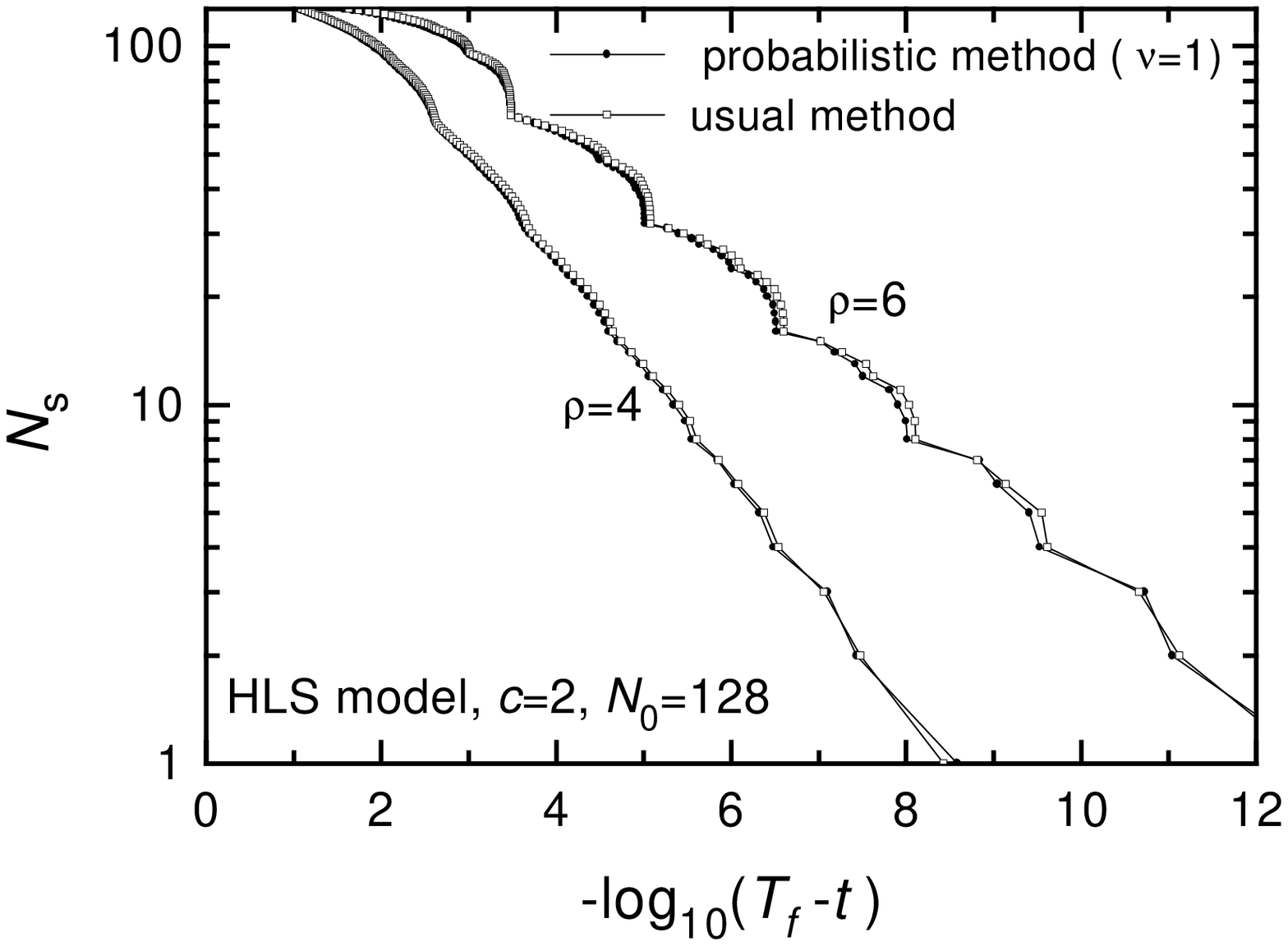}
\end{figure}


\begin{references}
\bibitem{herrman90} See, for example, {\sl Statistical Models for the Fracture of Disordered
Media\/}.
Editors, H.J. Herrman and S. Roux, North Holland (1990), and references therein.
\bibitem{coleman57} B.D. Coleman, Trans. Soc.\ Rheol. {\bf 1}, 153 (1957). J. Appl.\ Phys. {\bf
29}, 968 (1958)
\bibitem{smith81} P.L. Smith and S.L. Phoenix,
J. Appl.\ Mech.\ {\bf 103}, 75 (1981).
\bibitem{phoenix83} S.L. Phoenix and L.J. Tierney,
Eng. Fracture\ Mech.\ {\bf 18}, 193 (1983).
\bibitem{daniels45} H.E. Daniels,
Proc. Roy.\ Soc.\ {\bf A183}, 404 (1945).
\bibitem{harlow91} D.G. Harlow and S.L. Phoenix,
J. Mech. Phys.\ Solids\ {\bf 39}, 173 (1991).
\bibitem{turcotte82} D.L. Turcotte, R.F. Smalley and S.A. Solla,
Nature\ {\bf 313}, 671 (1985).
\bibitem{smalley85} R.F. Smalley, D.L. Turcotte and S.A. Solla,
J. Geophys.\ Res.\ {\bf 90}, 1894 (1985).
\bibitem{newman91} W.I. Newman and A.M. Gabrielov,
Int. J.\ Fracture\ {\bf 50}, 1 (1991).
\bibitem{Turcotte97} D.L. Turcotte. {\sl Fractals and Chaos in Geology and Geophysics\/}.
 2nd Edition. Cambridge University Press (1997).
\bibitem{Newman94} W.I. Newman, A.M. Gabrielov, T.A. Durand, S.L. Phoenix and D.L. Turcotte,
Physica D\ {\bf 77}, 200 (1994).
\bibitem{newman95} W.I. Newman, D.L. Turcotte and A.M. Gabrielov,
Phys. Rev. E\ {\bf 52}, 4827 (1995).
\bibitem{trece} The ELS case is an exception
because, for any \(j\), there exists only one sublist and hence all the
surviving elements decay at the same rate. This fact enables the formulation of
 the differential equation mentioned.
\bibitem{Gomez94} J.B. G\'omez, D.I\~niguez and A.F. Pacheco,
Phys. Rev.\ Lett.\ {\bf 71}, 380 (1993).
\bibitem{feynman48} R.P. Feynman,
Rev. Mod.\ Phys.\ {\bf 20}, 367 (1948).
\end{references}
\end{document}